\documentclass[a4paper,fleqn,usenatbib]{mnras}

\usepackage[T1]{fontenc}
\usepackage{ae,aecompl}

\usepackage{graphicx}	
\usepackage{amsmath}	
\usepackage{amssymb}	

\usepackage{newtxtext,newtxmath}

\newcommand\heiiline{HeII~$\lambda4686$}

\title[Outflows in Swift~J1357.2-0933]{Hot, dense HeII outflows during the 2017 outburst of the X-ray transient Swift~J1357.2-0933}

\author[Charles {\it et al.}]{Phil Charles,$^{1}$\thanks{E-mail: p.a.charles@soton.ac.uk}
James H. Matthews,$^{2}$
David A.H. Buckley,$^{3}$
Poshak Gandhi,$^{1}$
Enrico Kotze,$^{3, 4}$
\newauthor
and John Paice $^{1, 5}$
\\
$^{1}$Department of Physics \& Astronomy, University of Southampton, Southampton SO17 1BJ, UK\\
$^{2}$Astrophysics, Department of Physics, University of Oxford, Keble Road, Oxford OX1 3RH, UK\\
$^{3}$South African Astronomical Observatory, P.O.Box 9, Observatory, 7935, South Africa\\
$^{4}$Southern African Large Telescope, P.O.Box 9, Observatory, 7935, South Africa\\
$^{5}$Inter-University Centre for Astronomy \& Astrophysics, Post Box 4, Ganeshkhind, Pune-411007, India\\
}

\date{\today}

\pubyear{2019}

\begin{document}
\label{firstpage}
\pagerange{\pageref{firstpage}--\pageref{lastpage}}
\maketitle

\begin{abstract}
Time-resolved SALT spectra of the short-period, dipping X-ray transient, Swift~J1357.2-0933, during its 2017 outburst has revealed broad Balmer and HeII $\lambda$4686 {\it absorption} features, blue-shifted by $\sim$600 km s$^{-1}$.  Remarkably these features are also variable on the $\sim$500s dipping period, indicating their likely association with structure in the inner accretion disc. We interpret this as arising in a dense, hot ($\gtrsim30,000$K) outflowing wind seen at very high inclination, and draw comparisons with other accretion disc corona sources. 
We argue against previous distance estimates of $1.5$kpc and favour a value $\gtrsim6$kpc, implying an X-ray luminosity $L_X\gtrsim4\times10^{36}$ erg\,s$^{-1}$.  Hence it is {\it not} a very faint X-ray transient.
Our preliminary 1D Monte-Carlo radiative transfer and photoionization calculations support this interpretation, as they imply a high intrinsic $L_X$, a column density $N_H\gtrsim10^{24}$cm$^{-2}$ and a low covering factor for the wind. Our study shows that Swift~J1357.2-0933 is truly remarkable amongst the cohort of luminous, galactic X-ray binaries, showing the first example of \heiiline\ absorption, the first (and only) variable dip period and is possibly the first black hole `accretion disc corona' candidate. 
\end{abstract}




\section{Introduction}

The low-mass X-ray binary transients (XRTs) are short-period (usually $\leq$1d) interacting binaries, where a compact object (neutron star, NS, or black hole, BH) accretes material from its low-mass donor via an accretion disc.  Detected only during outburst we now know of $\sim$50 such objects, of which $\sim$75\% contain BHs, the remainder NSs, based on their X-ray and other properties.  These systems are outstanding laboratories for studying the late stages of stellar evolution, and determining fundamental properties of compact objects (see \citealt{charles_optical_2006}, \citealt{casares_mass_2014}, and the BlackCAT \citep{corral-santana_blackcat:_2016} and WATCHDOG \citep{tetarenko_watchdog:_2016} catalogues for more details).  The great majority of Galactic XRTs are in or close to the galactic plane, and only a handful appear to be at high ($\sim$70$^\circ$) inclination, and none are eclipsing. Furthermore, the last decade has seen an increase in those found at high latitude, one of which, Swift~J1357.2-0933 (hereafter J1357.2) is the subject of this paper.  

Discovered by Swift/BAT in 2011, J1357.2 peaked at a very low  L$_X\sim$10$^{35}$erg~s$^{-1}$, when compared to other BH XRTs, assuming a $\sim$1.5kpc) location \citep[][and references therein]{armas_padilla_swift_2014}.  This estimate was based on the optical identification of a substantially brightened ($\sim$6 mags) counterpart \citep{rau_swift_2011}, which had a very faint, red star at the same position on archival plates.  The outburst declined exponentially in X-rays, and was undetectable within 6 months \citep{Tetarenko18_Light-Curves}.  However, what drew attention to J1357.2 was the discovery \citep[][hereafter CS13]{corral-santana_black_2013} of periodic optical dipping.  Such dipping has been seen before in other short-period LMXBs \citep[e.g. A1916-05, see][] {Homer01_1916_dips}, and is due to the raised disc rim caused by the stream-impact from the donor, thereby producing obscuration of the inner disc regions when viewed at high inclination.  These dips therefore recur on the orbital period (usually $\sim$hours), but what was truly remarkable about J1357.2 was that, not only were the periodic dips recurring much too rapidly ($\sim$minutes) to be the binary period, but as the outburst progressed, the dip period was seen to lengthen.  Such variable-periodic dipping had never been seen before (nor since) in any other system.  It was interpreted by CS13 as being due to the gradual expansion of an inner torus in the disc, with associated structure (e.g. variable height) that led to the dipping.  At any given time, the dipping period would be that associated with the Keplerian rotation of the torus at that radius.

Optical spectroscopy of J1357.2 during outburst revealed strong, double-peaked H and HeI emission-lines, typical of a high-inclination disc.  Furthermore, CS13 found evidence for a 2.8$\pm$0.3h orbital period in their H$\alpha$ radial velocities, making it one of the shortest period systems amongst all the BH XRTs. This was refined by \cite{mata_sanchez_swift_2015} who inferred M$_1>$9.3M$_\odot$ and $q\sim$0.04, strengthening its BH candidature.  Such a short period would then imply a very low-mass donor, whose size CS13 suggested could be sufficiently small ($\leq$0.1R$_\odot$) that it would be permanently in the shadow of the accretion disc, thereby explaining the absence of X-ray or optical orbital modulations, even at a very high inclination.

Just 6 years after its discovery, J1357.2 produced a second outburst in 2017 \citep{drake_crts-ii_2017}, and the dips were again present and varying in frequency as during 2011 \citep[][hereafter P19]{paice_puzzling_2019}.  Furthermore, simultaneous SALT/NTT-ULTRACAM photometry and NuSTAR observations revealed no X-ray counterpart to the dips \citep[see also][]{beri_black_2019}, which unusually were shown to be blue, leading P19 to propose that they were caused by a moving, raised region within the inner disc that was obscuring a luminous, red emitting area.  In addition to P19's optical photometry, we also exploited the relative brightness of J1357.2 in outburst (V~16) to obtain SALT time-resolved spectra with the aim of searching for spectral signatures of the dips that could help explain their physical origin, and the resulting spectra are presented here.  Obtained later in the outburst, when the dipping period was $\sim$500s, these results should be viewed in parallel with those reported by \cite[][hereafter JI19]{JI19} taken a month earlier.

\section{Observations and Data Reduction}

We obtained medium-resolution spectroscopy of J1357.2 on 2017 April 28 (2$\times$1400s), 29 (4$\times$600s) and July 21 using the SALT Robert Stobie Spectrograph (RSS) and G0900 grating (Buckley et al. 2006).  With the July 21 spectroscopy our aim was to temporally resolve the absorption dips, so we used RSS in Frame-Transfer mode, obtaining 31 100s exposures (with no dead-time) during the SALT track.  The dip period at this time was $\sim$500s (see P19).  Combined with a 1.5 arcsec slit, all spectra covered the $\lambda\lambda$4100-7100 range at a resolving power of $\sim$1000.

\begin{figure}
	\includegraphics[width=\columnwidth]{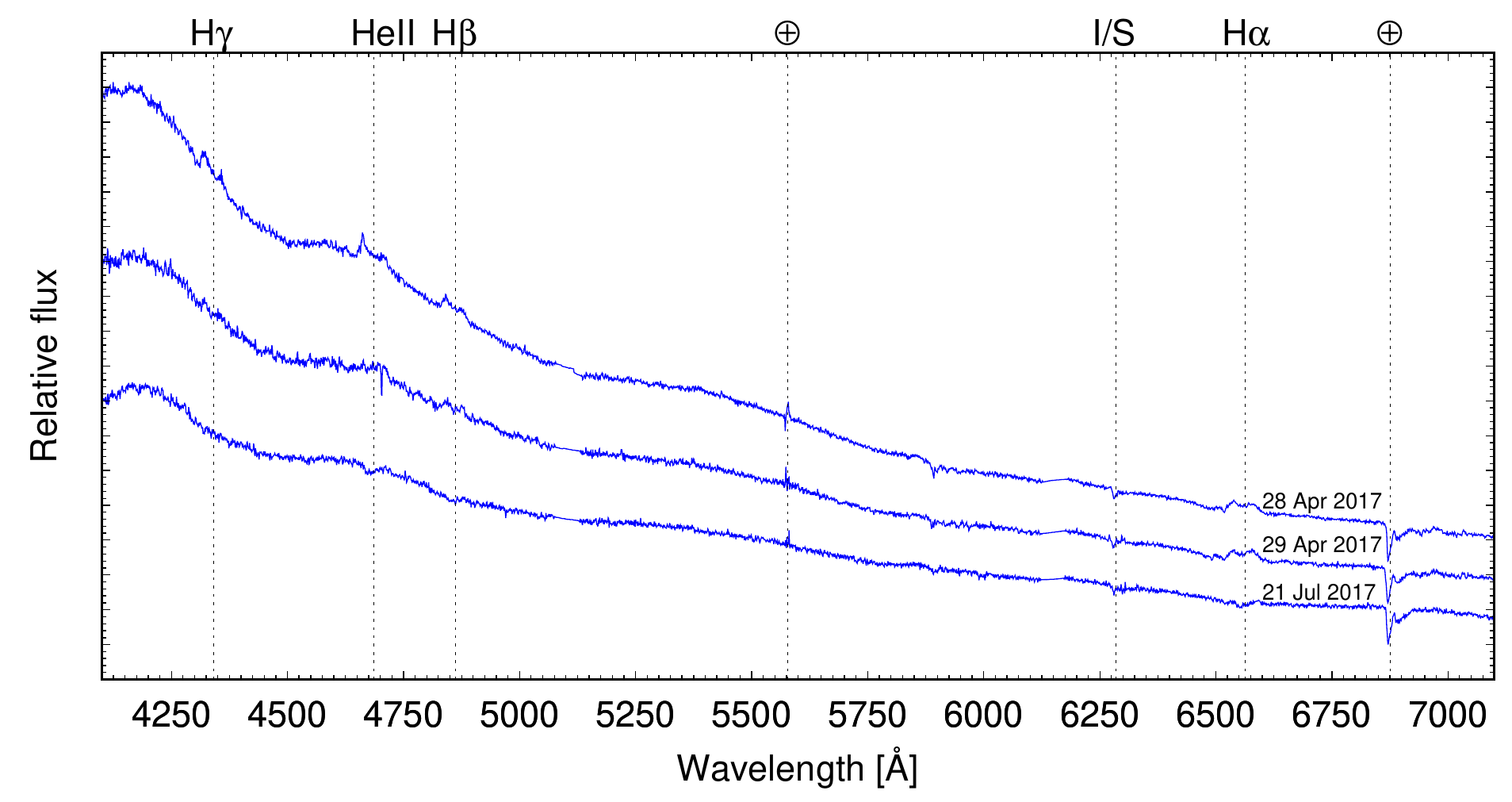}
    \caption{Mean SALT RSS spectra of J1357.2, with key spectral features marked with dotted lines, including the diffuse interstellar band at $\lambda$6283.  The April spectra are similar to those of CS13 with double-peaked HeII and Balmer emission profiles typical of a high-inclination disc.  Terrestrial atmospheric features (OI, B-band) are marked $\oplus$.}
    \label{fig:SALT_mean}
\end{figure}

These data, plus associated arcs and bias frames, were pipeline-processed  \citep{crawford_pysalt:_2010}, followed by cosmic-ray removal using the \texttt{IRAF} task L.A. Cosmic \citep[][]{2001PASP..113.1420V} on the 2D image, and wavelength calibrated using standard \texttt{IRAF} \citep{tody_iraf_1986}
routines.  While SALT's continuously varying entrance pupil precludes accurate flux calibration, these spectra have been divided by a suitable standard star \citep[LTT4364:][]{1994PASP..106..566H} to remove instrumental response features.  The July time-series average spectrum is shown in Figure \ref{fig:SALT_mean}, along with Apr 28/29 for comparison.

\begin{figure}
	\includegraphics[width=\columnwidth]{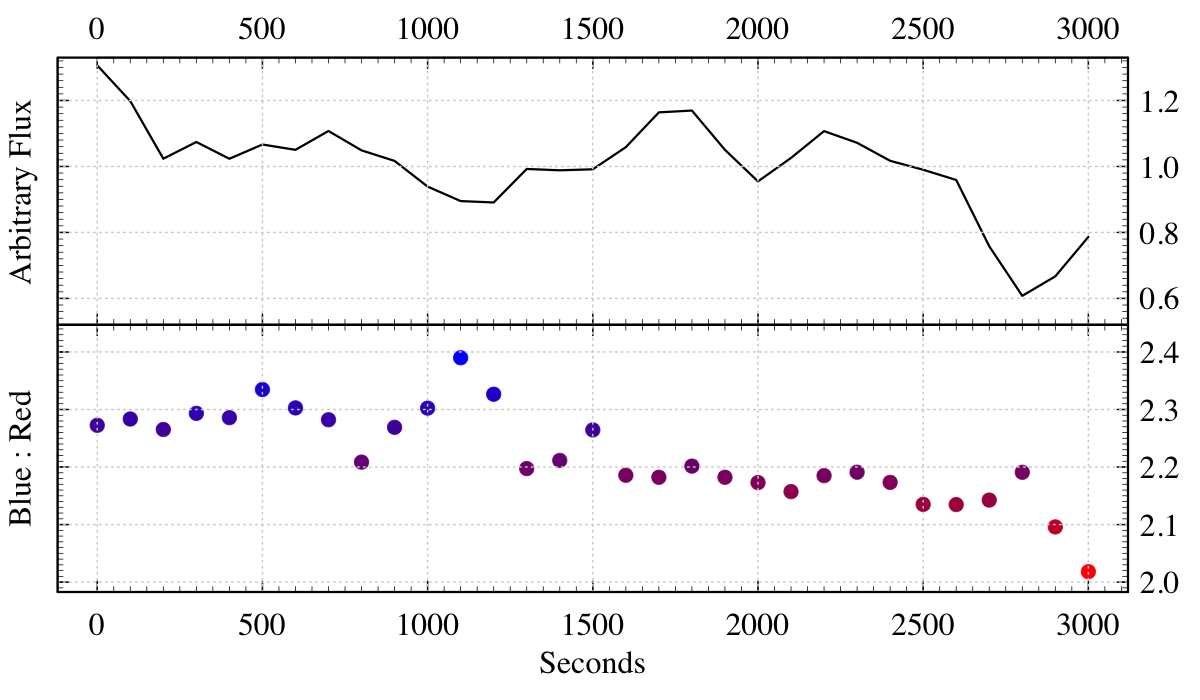}
    \caption{{\it Upper:} Light curve of J1357.2 (relative to its mean) from the July 21 RSS spectra by using a comparison star on the slit. {\it Lower:} Approximate blue/red colour variations from the total flux in the blue ($\lambda\lambda$4100-5100) and red ($\lambda\lambda$6200-7100) CCDs, relative to the same star.}
    \label{fig:SALT_spec_lc}
\end{figure}

An indication of the variability present during the July SALT run (Figure \ref{fig:SALT_spec_lc}) uses a nearby field star (SDSS J135716.43-093140.1) also on the slit for reference.  Colour variations are shown by using the flux ratio in the blue and red RSS CCDs.  There is clear variability on the dip timescale ($\sim$500-800s), and we note that the spectrum at time 1100s is particularly blue.

\begin{figure}
	\includegraphics[width=\columnwidth]{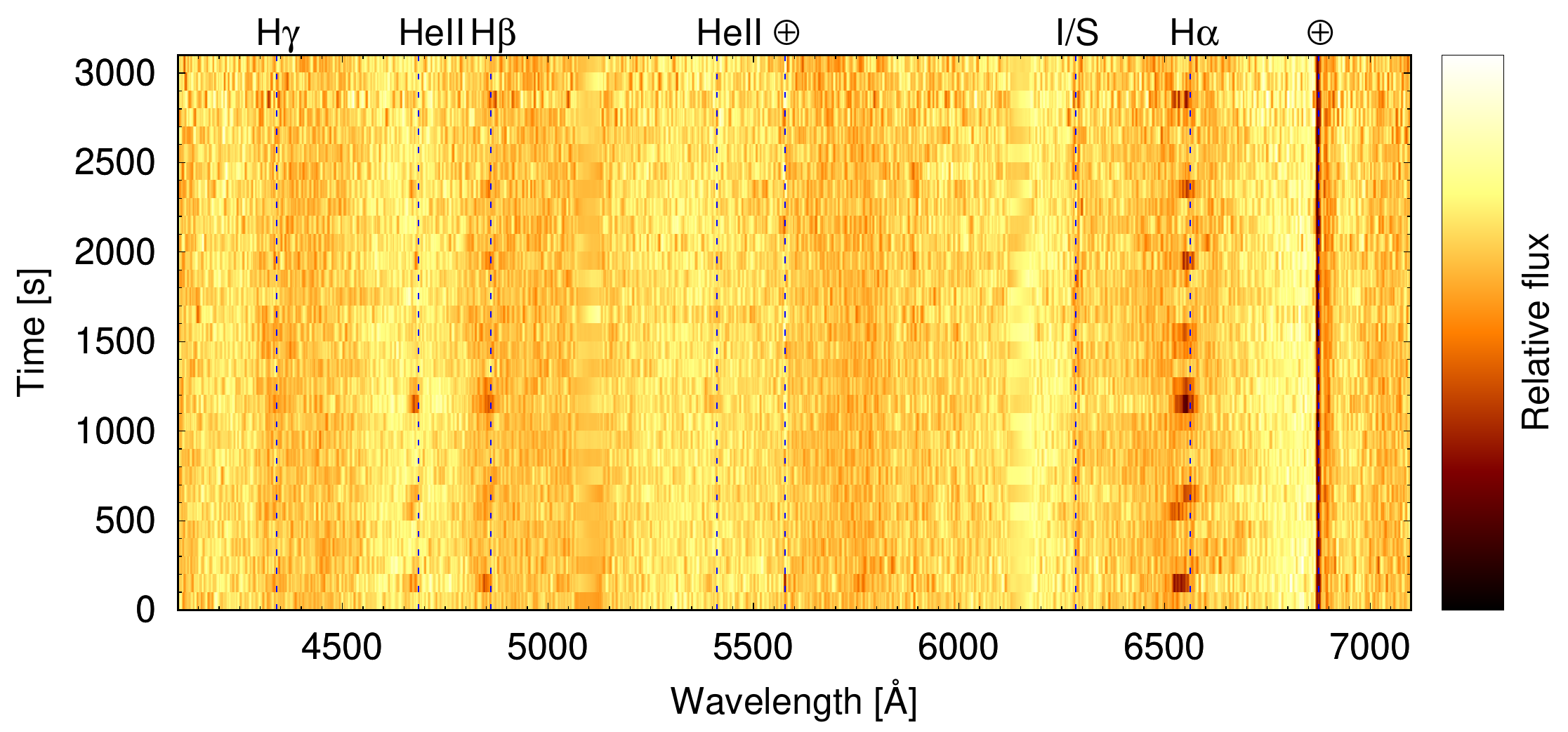}
    \caption{SALT time-series spectra of J1357.2 on 21 July 2017, displayed as a grey-scale with time increasing from bottom to top.  Key spectral features are marked with dotted lines. Note in particular the strong variations in \heiiline\ and H$\beta$ that are associated with the dipping frequency.}
    \label{fig:SALT_spec_time_series}
\end{figure}

\begin{figure}
    \includegraphics[width=\columnwidth]{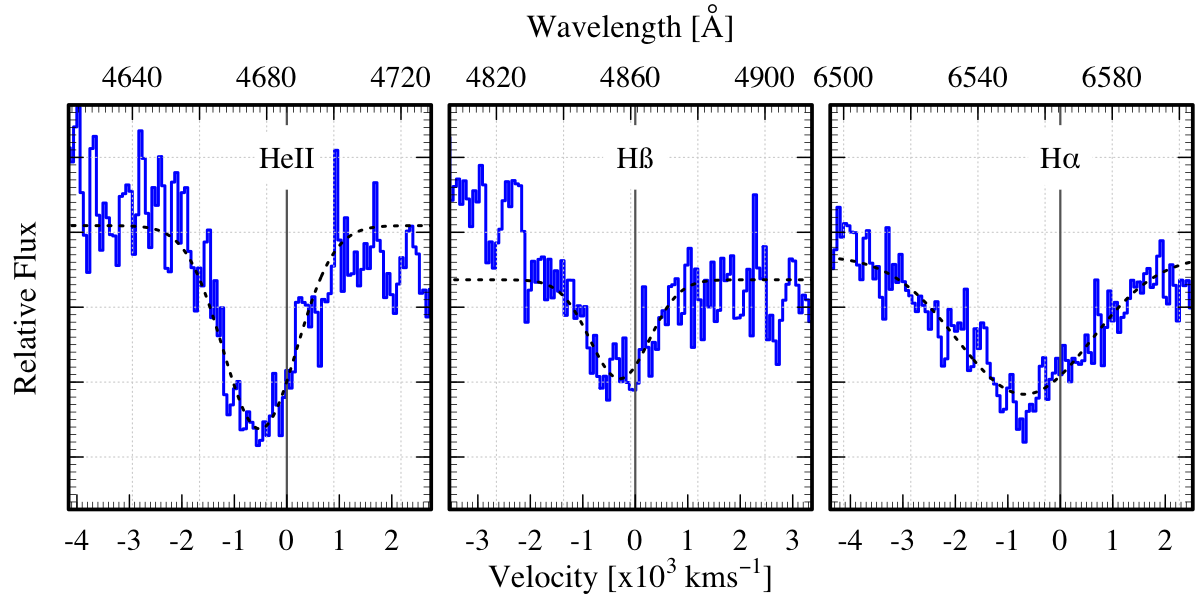}
    \caption{Zoomed sections of the mean of four SALT RSS spectra of J1357.2 on 2017 July 21, where \heiiline, H$\beta$ and H$\alpha$ absorption features are strongest. Gaussian fits to the line profiles are shown, as are the velocities relative to rest wavelength.}
    \label{fig:SALT_blue_dip}
\end{figure}

The key spectral feature that is highly variable is the broad, blue-shifted ($\sim$600 km s$^{-1}$) HeII $\lambda$4686 and Balmer absorption.  These variations are highlighted in Fig. \ref{fig:SALT_spec_time_series} which displays the 31 individual spectra as a grey-scale.  The features are strongest in spectra 7, 12, 13 and 24, see Figure \ref{fig:SALT_blue_dip} zoomed-in on HeII/H$\beta$ and H$\alpha$.  Fitting single Gaussians gives centroid blue-shifts of 8$\pm$1.5 \AA\ for HeII and H$\alpha$, corresponding to $\approx600$km~s$^{-1}$, and FWHM in the range 500-700 km s$^{-1}$. The maximum blueshift is around $1500$km~s$^{-1}$ in \heiiline\ and H$\beta$, and even higher in H$\alpha$.

\section{Modelling an Unusual Disc Wind}

Such a large blue-shift in J1357.2's absorption features, also seen by JI19 a month earlier, is a {\it smoking gun} signal of outflowing material, presumably a disc wind.  Spectral signatures (usually in the form of blue-shifted broad absorption lines [BALs] or P-Cyg profiles) of outflowing material from accreting objects have regularly been seen at all wavelengths.  Indeed, they appear ubiquitous, having been well-studied across the full mass spectrum, from cataclysmic variables (CVs) to supermassive BHs \citep[see e.g.][for a review]{matthews_disc_2016}. In X-ray binaries, equatorial disc winds are thought to be produced in the soft state as inferred from broad Fe \textsc{xxv}/\textsc{xxvi} absorption \citep{ponti_ubiquitous_2012}. In addition, blue-shifted HeI and H$\alpha$ absorption has been detected in V404 Cyg \citep{casares1991,munoz-darias_regulation_2016,munoz-darias_2017,mata_sanchez_1989_2018}, V4641 Sgr \citep{munoz-darias_low-luminosity_2018}, MAXI J1820+070 \citep{munoz-darias_2019} and CVs \citep{kafka_detecting_2004}. Optically detected P-Cygni profiles indicate the presence of cooler ($\sim10,000$K) outflowing material compared to the X-ray wind detections, though the extent to which the two phenomena are related is not yet known.

\subsection{Radiative Transfer Method and Model Setup}
Modelling a disc wind in J1357.2 is challenging given the lack of constraining information about the outflow geometry and intrinsic radiation source. We assume that the outflow is photoionized by a central X-ray source and all heating sources are radiative, which may or may not be reasonable. For our modelling, we used a Monte Carlo radiative transfer code, confusingly known as \textsc{python} (see  \cite{long_modeling_2002}, with subsequent updates and testing described by \cite{sim_two-dimensional_2005}, \cite{higginbottom_simple_2013} and \cite{matthews_impact_2015}). The code first calculates the outflow's ionization and temperature structure, before computing the detailed spectrum from the converged model. We adopt the hybrid macro-atom approach to line transfer, described by \cite{matthews_impact_2015}, in which H and He are treated as macro-atoms according to the formalism of \cite{lucy_monte_2002,lucy_monte_2003}, whereas metals are treated as `simple-atoms'. This is a well-tested technique that has been used to model line formation when there are large departures from LTE, such as in supernovae \citep[e.g][]{kromer_time-dependent_2009}. Convergence criteria are described by \cite{long_modeling_2002} and, briefly, require that the electron temperature, $T_e$, has stopped changing significantly and that heating and cooling are closely balanced. The model spectra presented here come from models with $100\%$ convergence, although some of the models (12/54) in the simulation grid did not converge well. The poorly converged models were all at either the extreme low or high ionization end of the simulating grid, and the models in between allowed us to cover the region of parameter space where \heiiline\ first appeared and then disappeared. Nevertheless, further exploration of parameter space is merited.  

Our 1D spherically-symmetric model uses 30 radial cells separated logarithmically \cite[see][]{matthews_disc_2016}. The velocity at radius $R$ is set according to a so-called $\beta$-law, $v(R) = v_0 + (v_\infty-v_0) \left[ 1 - (R_0/R) \right]^\beta$, where $R_0$ is the initial radius, $v_0$ and $v_\infty$ the initial and terminal velocities and $\beta$ governs the rate of acceleration. This velocity law is commonly used to model stellar winds \citep[e.g.][]{puls_o-star_1996}, and, while it may not be correct, it gives an appropriate starting point. The density, $\rho(R)$, is calculated from the isotropic mass loss rate, $\dot{M}_{\mathrm{w,iso}} = 4\pi R^2 \rho v$. The illuminating spectral energy distribution (SED) is $F_\nu \propto \nu^{-\alpha}$, with a low-frequency cutoff at $\nu_c=10^{17}$Hz and 2-10 keV luminosity $L_{2-10}$. We conducted a small grid of simulations over the parameters $L_{2-10}$ (from $10^{35}-10^{37.5}$erg~s$^{-1}$ in steps of $0.5$~dex), $\dot{M}_{\mathrm{w,iso}}$ ($ 10^{-5},10^{-6},10^{-7}~M_\odot~\mathrm{yr}^{-1}$) and $\beta$ (0.5, 1, 4).  We fixed $\alpha=0.8$, $v_0=100$km~s$^{-1}$, $v_\infty=3000$km~s$^{-1}$ and set $R_0$ to the approximate radius at which the Keplerian period equals the dip period ($2\times10^{10}$cm). Note that $\dot{M}_{\mathrm{w,iso}}$ is not the true mass-loss rate, rather a normalisation of the flow density. The true rate will be significantly lower due to the covering factor of the wind. 

\begin{figure}
	\includegraphics[width=1.0\columnwidth]{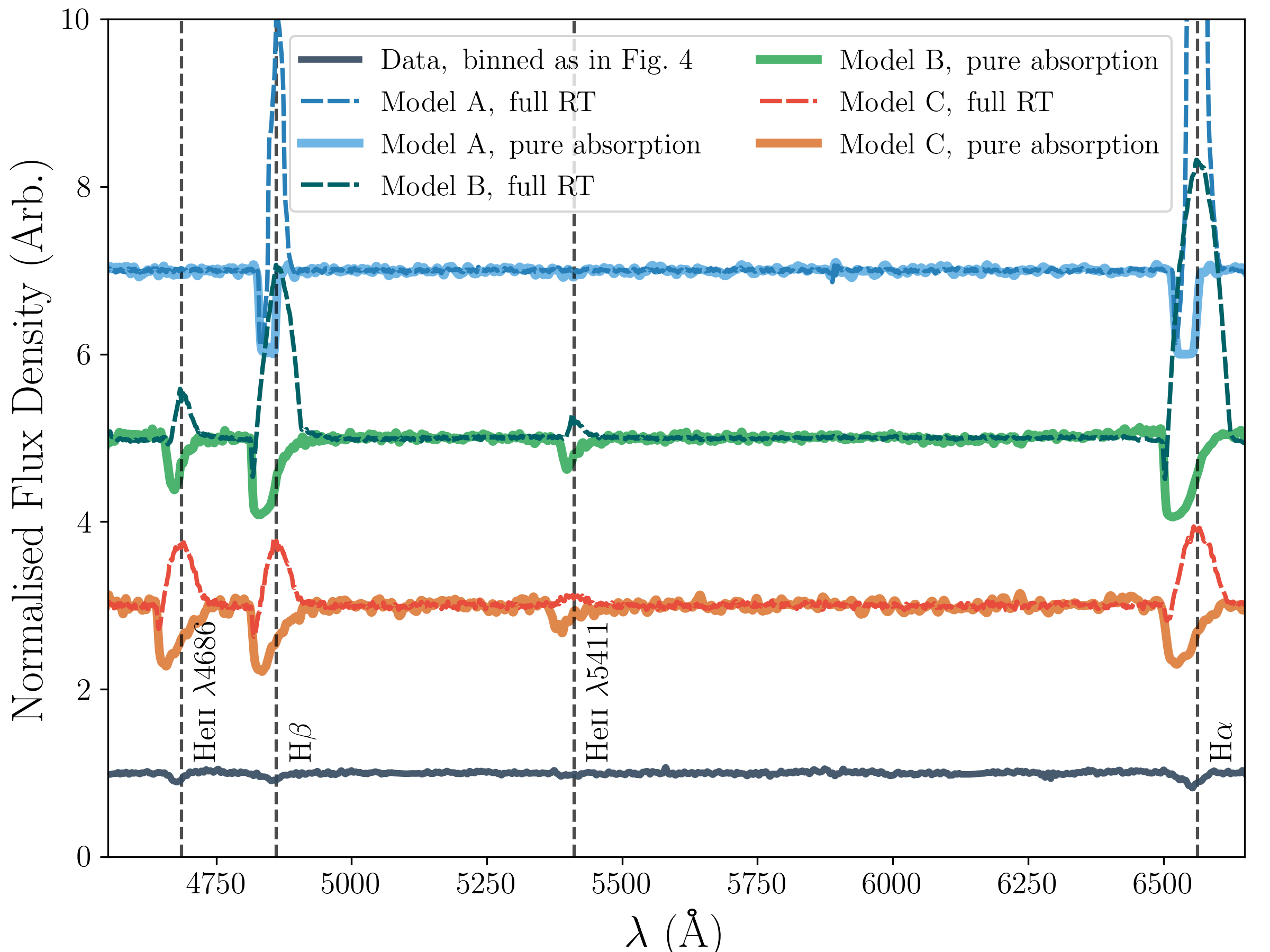}
    \caption{Continuum normalised model spectra, calculated with full radiative transfer and in the pure absorption limit, compared to observational data. We show three models that produced obvious \heiiline\ and/or Balmer absorption lines: Model A ($\dot{M}_{w,\mathrm{iso}}=10^{-7}M_\odot~\mathrm{yr}^{-1}, L_{2-10}=10^{35}~\mathrm{erg~s}^{-1}, \beta=4$), Model B ($\dot{M}_{w,\mathrm{iso}}=10^{-6}M_\odot~\mathrm{yr}^{-1}, L_{2-10}=10^{36}~\mathrm{erg~s}^{-1}, \beta=0.5$), and Model C ($\dot{M}_{w,\mathrm{iso}}=10^{-5}M_\odot~\mathrm{yr}^{-1}, L_{2-10}=10^{37}~\mathrm{erg~s}^{-1}, \beta=0.5$). See text for details. For clarity, offsets of $+6$,$+4$ and $+2$ are applied to models A, B and C, respectively. }
    \label{fig:model_spectra}
\end{figure}

\subsection{Simulation Results}
We find that a fraction (9 out of 54) of our models produce \heiiline\ absorption, appearing as P-Cygni profiles due to the model's spherical symmetry. Fig.~\ref{fig:model_spectra} compares the data from Figure \ref{fig:SALT_blue_dip} with two spherical models from our simulations. For each model, we include the spectrum as calculated with full radiative transfer as well as in the pure absorption limit, where radiation packets are discarded once macro-atoms are activated. In the pure absorption limit, the models show clear blue-shifted BALs in H$\alpha$, H$\beta$ and \heiiline. In the full RT case, these absorption profiles are mostly swamped by emission, and, notably, the emission and absorption profiles do not sum to zero as would be expected from pure scattering lines in spherical symmetry. This indicates that there is a non-scattering contribution to the line source function. One interpretation of the pure absorption spectrum observed is that it is due to a low covering factor for the wind. The pure absorption model spectrum also shows much deeper absorption troughs compared to that observed, which could be due to partial covering of the optical continuum or emission lines from the wind partially filling in the troughs.  The absorption troughs in both the models and the data show absorption redward of line centre. In the data, this might be caused by rotation. In the models, it is due to obscuration of the reprocessed continuum by the far side of the wind, since $\nu_c=10^{17}$Hz and the optical continuum is entirely produced by X-ray reprocessing. In models where \heiiline\ absorption is produced, this reprocessed continuum actually peaks in the UV and the soft X-rays are heavily absorbed by the wind, behaviour which is consistent with the findings of \cite{beri_black_2019} from UV and X-ray observations during the same outburst.

In our simulation grid, we required the following properties to produce \heiiline\ absorption:
(i) $L_{2-10}\gtrsim10^{36}$ erg s$^{-1}$; (ii) $\dot{M}_{\mathrm{w,iso}} \gtrsim 10^{-6}~M_\odot~\mathrm{yr}^{-1}$, leading to maximum densities of $n_H\sim10^{13-14}$~cm$^{-3}$ and column densities of $N_H\gtrsim10^{24}$~cm$^{-2}$; (iii) corresponding ionization parameters of $\xi\sim100$ and $T_e\approx20,000-70,000$K. We found that models with lower $L_{2-10}$ were under-ionized and only produced Balmer absorption or none at all (an example is shown in Fig. ~\ref{fig:model_spectra}). Models with lower mass loss-rates did not have sufficient line opacity. The modelling results give us clues to the physical conditions in the outflow, but they should be interpreted with caution. For example, we may be limited by our imposed velocity law or geometry. In particular, in a 1D model there is only one route for X-rays to reach the outer parts of the wind, whereas we already know the X-ray source is not eclipsed and the absorption lines are phase dependent, which means that 2D and 3D radiative transfer effects {\em must} play a role. We therefore cannot draw robust constraints on parameters from this modelling. However, importantly, we have demonstrated that \heiiline\ absorption can be produced for reasonable system parameters and is suggestive of a relatively high intrinsic $L_X$, as well as a high density and a low covering factor for the outflow.

\section{Discussion}
Our time-resolved spectroscopy of J1357.2 has revealed a hot spectral component that appears associated with the optical absorption dips.  What is remarkable about this component is that it has \heiiline\ in absorption, a feature normally only seen in very hot stars (early O stars, \citealt{walborn_erratum:_1990}, and DO/DB white dwarfs, \citealt{wesemael_atlas_1993}). Yet the substantial blue-shift in all the strong absorption features is compelling evidence for associating them with an accretion disc wind, as modelled in the previous section, launched from a region which gradually moves out through the disc as the outburst progresses.  Indeed we note that our observed blue-shift is significantly less than that seen by JI19, as expected given the lower escape velocity at the larger radius. \heiiline\ absorption has never been seen in a disc wind before, although there is evidence in \cite{quimby_sn_2007} and \cite{dessart_using_2008} for its presence during the shock break-out phase in Type IIP supernovae, while He\textsc{ii}~$\lambda1640$ wind features have been seen in FUV spectra of CVs \citep{long1991,hoare_he_1994}.

Amongst the now several hundred LMXBs known, J1357.2 has the distinction of exhibiting two quite unique properties never seen before -- variable frequency dipping and a hot disc wind.  We therefore re-examine the fundamentals of J1357.2:

\begin{enumerate}
    \item {\bf Orbital period.}  This was determined by CS13 from their outburst spectroscopy to be 2.8$\pm$0.3hr, using H$\alpha$ emission radial velocities, making it one of the two shortest-period BH XRTs known.  Normally we would expect the period to be confirmed in other ways, usually via optical or X-ray light-curves, especially if the inclination is thought to be high.  However, as yet, no additional periodic variation has been seen at any wavelength, save the fast optical absorption dips (and seen only once in X-rays, AP14a).  Nevertheless, the H$\alpha$ double-peak separation of 1790 km s$^{-1}$ led CS13 to use the \cite{orosz_orbital_1995} relation between outer disc velocity and $K_2$ to infer $K_2\geq$690 km s$^{-1}$.  Furthermore, this is independently supported by the \cite{casares_fwhm-k2_2015} relation between H$\alpha$ FWHM and $K_2$ which gives $K_2=$760 km s$^{-1}$, so these values are all self-consistent with those expected for a very low-mass donor orbiting a BH with a 2.8 hr period and indicate $f(M)>6M_\odot$. But a direct observational confirmation of this period is still highly desirable.
    
    \item {\bf Distance.}  The presence on the DSS of a faint ($r\sim$22), very red star, led \cite{rau_swift_2011} to identify this as the quiescent donor, an $\sim$M4 star at $d\sim$1.5kpc.
    Such a relatively close location was not unexpected given its very high ($b\sim$50$^\circ$) galactic latitude, but this implied only $L_X\sim$10$^{35}$ erg s$^{-1}$ at outburst peak, making it a member of the class of ``very faint X-ray transients'' (or VFXTs).  Its X-ray spectral properties (AP13) were consistent, as it remained in the hard state at maximum, but J1357.2 would then be the only BH VFXT!  Nevertheless, once in quiescence, even at the DSS magnitude, it was an attractive target for VLT spectroscopy, as the short $P_{orb}$ and BH candidature imply a high $K_2$ which should be straightforward to measure.  This was attempted by \cite{torres_vlt_2015}, but their spectrum was essentially identical to that seen in outburst, revealing no donor star features. More recently, the $\sim$6-yr OIR light-curve of J1357.2 \citep{russell_optical_2018} showed sustained strong variability over the V$\sim$20-22 range outside outburst, indicative of a substantial and continuing disc-emitting component. Accordingly, with no evidence for the donor on archival surveys, support for the 1.5kpc distance disappears. This was also discussed by \cite{shahbaz_evidence_2013}, who used the outburst amplitude - $P_{orb}$ relation to extend the possible distance range out to $\geq$6.3kpc, thereby increasing all the luminosity estimates by $\sim\times$40 to at least $L_X \sim 4\times10^{36}$erg~s$^{-1}$, and hence it is {\it not} a VFXT.
    
    \item {\bf Inclination.}  While a range of inclinations has been considered for J1357.2, (see also P19), we believe the presence of the variable frequency optical dips argues very strongly for a high inclination.  Both CS13 and AP14a interpret J1357.2 as an Accretion Disc Corona (ADC) source, in which $i$ is so high ($>$80$^\circ$) that the central X-ray-emitting region is not directly visible.  Instead, only X-rays scattered from the surrounding corona are observable, inferring an intrinsic $L_X$ at least two orders of magnitude greater (see \citealt{charles_lmxbs:_2011} and references therein).  It had already been noted that the X-ray to optical flux ratio of J1357.2 was low, and this is confirmed by comparison with the similar short-period, high-latitude XRT, Swift~J1753.5-0127.  Simultaneous X-ray/optical flux data were taken of both in outburst by Swift, giving a de-reddenned $F_X/F_{UVopt}\sim$0.3 for J1357.2 \citep{beri_black_2019}, but $\sim$150 for Swift~J1753.5-0127 \citep{shaw_ZCam19}, i.e. $\sim$500 times greater, as expected if J1357.2 is an ADC source. Curiously, the well-known ADC sources (such as 2A1822-371, M15/AC211 and 4U2129+47) are demonstrably NS systems, which would make J1357.2 the first BH example.  The key argument against the ADC interpretation is that all established ADC sources show extensive partial optical and X-ray eclipses, whereas J1357.2 shows none.  However, CS13 pointed out that the very short $P_{orb}$ of J1357.2 would require a donor mass $\leq$0.2 $M_\odot$ which would be entirely shadowed by the disc rim, and therefore unable to produce any X-ray eclipse.  The donor would of course eclipse the disc itself, but the disc variability is so great as to likely render the eclipse unobservable.   Furthermore, \cite{mata_sanchez_swift_2015} point out that the deep absorption core in HeI $\lambda$5876 also indicates a high $i$ ($>$75$^\circ$ when observed in CVs).

\end{enumerate}

\subsection{What drives the outflow?} 
Disc winds can be driven by thermal pressure, radiation pressure or magnetic fields \cite[e.g.][]{proga2007}. It is difficult to draw robust conclusions about the driving mechanism in J1357.2, but we provide a brief discussion. We adopt a BH mass of $10~M_\odot$ and $L_\mathrm{Edd} \sim 10^{39}$erg~s$^{-1}$, typical values for BH XRTs \citep[see e.g.][]{ozel_black_2010}. 

Thermal winds are caused by the X-ray corona heating the disc surface, causing the sound-speed in the irradiated disc to exceed the escape velocity \citep{begelman_compton_1983}. It has been suggested that the equatorial disc winds observed in X-ray binaries such as GRO J1655-40 and H1743-322 can be explained by thermal and thermal-radiative driving \citep{higginbottom_radiation-hydrodynamic_2018,done_thermal_2018,tomaru_thermal-radiative_2019}. The Compton radius $R_C \approx 2\times 10^{11}~T_C^{-1}~\mathrm{cm}$, where $T_C$ is the Compton temperature in units of $10^8$K, which is comparable to the Compton temperature for an $\alpha=0.8$ power-law. The critical luminosity is $L_\mathrm{crit} \approx 0.03~T_C^{-1/2}~L_\mathrm{Edd}$. The association of the outflow with the dipping period suggests a launch radius of $\sim2\times10^{10}$cm, less than $R_C$ but close to the critical value of $0.2~R_C$ discussed by, e.g., \cite{done_thermal_2018}. Thus thermal winds may just be feasible if $L_X\gtrsim10^{37}$~erg~s$^{-1}$. However, the outflowing material in thermal winds tends to stay close to $T_{C}$, whereas we expect the He~II absorbing material to have temperatures $\sim40,000$K. We note that our inferred launch radius has an escape velocity comparable to our observed terminal velocity (see also JI19).

Radiative driving can be mediated by different opacity sources. At luminosities significantly below $L_{Edd}$, additional opacity sources involving bound-free or line transitions are needed to provide the necessary force. The effect is normally quantified in terms of the `force multiplier', ${\cal M}$, the factor by which the radiation force is enhanced above the Thomson scattering value. Given their hard spectra relative to AGN and CVs, line-driven winds are not usually invoked in X-ray binaries. However, \cite{tomaru_thermal-radiative_2019} have shown that line and bound-free opacity might supplement thermal driving, while photoionization calculations \citep{dannen_photoionization_2018} produce ${\cal M}\sim 10$ for $\xi\sim100$ plasmas with significant He~II opacity. Comparison with $L_{Edd}$ suggests that radiative driving might be possible for $L_\mathrm{bol}\sim10^{38}$erg~s$^{-1}$. Such a `boost' from line-driving seems reasonable given the presence of absorption lines, but it still requires an extremely high intrinsic luminosity. Given the apparent difficulties with radiative and thermal driving mechanisms, it is tempting to invoke driving from either magneto-centrifugal effects \citep{blandford_hydromagnetic_1982} or magnetic pressure \citep[e.g.][]{stone_norman_1994}. Alternatively, if the luminosity is indeed quite low, the outflow may originate in a radiatively inefficient, low accretion rate flow \citep{blandford1999}. At this stage, {\em known} driving mechanisms (thermal or radiative) are only feasible for an intrinsic luminosity higher than that implied for a $d\sim6$kpc. A higher intrinsic luminosity might therefore be explained by an even larger distance and/or an extremely high inclination, and could bring the X-ray to optical flux ratio in line with canonical values for X-ray binaries. However, the source still remains puzzling in terms of its location in the radio-X-ray plane. \cite{Plotkin16_sightline} showed the source is radio sub-luminous relative to other BH binaries, and our proposed scenario reinforces this.


\vspace{-2em}
\section*{Acknowledgements}
 We thank the anonymous referee for their constructive comments. This work is part of the SALT Large Programme 2016-2-LSP-001 (PI: Buckley).  DAHB acknowledges support from the National Research Foundation. We thank T. Munoz-Darias and F. Jim\'{e}nez Ibarra for sharing their GTC results. We also thank R. Fender, S. Motta, J. Bright, N. Degenaar, S. Sim, K. Long, C. Knigge and N. Higginbottom for discussions. PAC acknowledges the Leverhulme Trust for an Emeritus Fellowship. JM is funded by STFC grant ST/N000919/1. We gratefully acknowledge the use of matplotlib \citep{Hunter2007}. PG and JAP thank STFC and a UGC/UKIERI Thematic Partnership.

\vspace{-1em}



\bibliographystyle{mnras}
\bibliography{J1357_outflows}




\bsp	
\label{lastpage}
\end{document}